\documentclass[a4paper,11pt]{article}
\pdfoutput=1 

\usepackage{jinstpub} 

\title{\boldmath VOXES: a high precision X-ray spectrometer for diffused sources with HAPG crystals in
the 2-20 keV range}

\author[a,1]{A. Scordo,\note{First author.}}
\author[a]{C. Curceanu,}
\author[a]{M. Miliucci,}
\author[b,a]{H. Shi,}
\author[a,c]{F. Sirghi}
\author[d]{and J. Zmeskal}

\affiliation[a]{INFN Laboratori Nazionali di Frascati, \\Frascati (Roma), Italy}
\affiliation[b]{HEPHY - Institut f\"ur Hochenergiephysik der \"OAW, \\Vienna, Austria}
\affiliation[c]{Horia Hulubei National Institute of Physics and Nuclear Engineering (IFIN-HH), \\Magurele, Romania}
\affiliation[d]{Stefan-Meyer-Institut f\"ur subatomare Physik, \\Vienna, Austria}

\emailAdd{alessandro.scordo@lnf.infn.it}

\abstract{Bragg spectroscopy is one of the best established experimental methods for high energy
resolution X-ray measurements and has been widely used in several fields,
going from fundamental physics to quantum mechanics tests, synchrotron radiation and X-FEL applications, 
astronomy, medicine and industry. However, this technique is limited to the measurement of photons
produced from well collimated or point-like sources and becomes quite inefficient
for photons coming from extended and diffused sources like those, for example, emitted in the exotic atoms radiative transitions.
The VOXES project's goal is to realise a prototype of a high resolution and
high precision X-ray spectrometer, using Highly Annealed Pyrolitic Graphite 
(HAPG) crystals in the Von Hamos configuration, working also for extended sources.
The aim is to deliver a cost effective system having an energy resolution at the
level of eV for X-ray energies from about 2 keV up to tens of keV, able to perform 
sub-eV precision measurements with non point-like sources.
In this paper, the working principle of VOXES, together with first results, are presented.}


\proceeding{24$^{\text{th}}$ International Congress on X-ray Optics and Microanalysis\\
  25-29 September 2017\\
  Trieste}

\begin{document}
\maketitle
\flushbottom

\section{Introduction}
\label{sec:intro}

High precision measurements of X-rays are important in many fields of fundamental science, from particle and nuclear
physics to quantum mechanics, as well as in astronomy and in several applications using synchrotron
light sources or X-FEL beams, in biology, medicine and industry. 
\\An intensive effort has been pursued 
in the last years to optimize and develop large area X-ray detection systems;
new types of detectors are being constantly realized, with improved performances in terms of
efficiency, energy resolution and costs. 
\\ For spectroscopy applications, among the commercial available solid state devices, the best performances in terms of 
spectral resolution are provided by Silicon Drift Detectors (SDDs), recently used as large area spectroscopic detectors 
in the framework of the SIDDHARTA experiment \cite{SIDD} for exotic atoms transition lines measurements. 
\\However, when few eV linewidths have to be measured or precisions below $1\,eV$ are needed, these devices are limited by their 
intrinsic resolution of about 120 eV (FWHM) at 6 keV, related to the Fano Factor and to the electronic noise.
Experiments performed in the past at the Paul Scherrer Institute (PSI), measuring pionic atoms \cite{PIAT}, 
pioneered the possibility to combine Charged Coupled Device detectors (CCDs) with crystals, using Bragg reflection principle. 
The used crystals were of silicon type and the energy range achievable to the system was limited to few keV, 
due to the crystal structure, which is also responsible for the low intrinsic efficiency of these type of crystals.
\\Recently, a new type of high resolution detectors are under development: the Transition Edge
Sensors (TES)\cite{TES1}. Their achievable energy resolution is excellent (few eV at 6 keV), but their usage is still limited
by a very small active area and the prohibitively high costs of the complex cryogenic system needed 
to reach the operational temperature of $\simeq\,50\,mK$, which also makes their use rather laborious.
\\In spite of the fast evolving radiation detectors market, the aim of performing sub-eV precision measurements 
not only of photons produced from well collimated and point-like sources
but also from extended and diffused sources, like those used to study the exotic atoms radiative transitions, 
is still to be reached.

\section{Von Hamos spectrometer with HAPG mosaic crystals}
\label{sec:vh-hapg}

A new possibility to reconsider Bragg spectrometers as good candidates for measuring diffused X-rays comes from the
development, in the last decades, of the Higly Annealed Pyrolitic Graphite mosaic crystals (HAPG\cite{HAPG}),
consisting in a large number of nearly perfect small pyrolitic graphite crystallites, randomly misoriented around a preferred direction. 
The FWHM of this random angular distribution, called mosaicity, together with a lattice spacing constant of $3,514$ {\AA}, 
enables them to be highly efficient in diffraction in the 2-20 keV energy range for the n=1 reflection order. 
Higher energies can be reached for higher orders. 
\\The production mechanism of HAPG allows to obtain crystals with different ad-hoc geometries; this feature makes them suitable
to be used in the Von Hamos configuration \cite{VON1}, combining the dispersion of a flat crystal with the focusing properties of cilindrically
bent crystals. 
\\Von Hamos spectrometers have been extensively used in the past proving very promising results in terms of spectral resolution;
for example, one of the first measurements, performed in 2002 by Shevelko et al \cite{HAP1}, reported $E / \Delta E$ values ranging from 
800 (Mg, first diffraction order) to 2000 (Fe and Ti for higher diffraction orders) using a $200 \mu m$ thick curved mica crystal with a $20\,mm$ 
radius of curvature. More recently, measurements performed in 2013 by Zastrau et al \cite{HAP2} reported similar results with $100\,\mu m$ and $40\, \mu m$
thick curved HOPG (Highly Oriented Pyrolitic Graphite) crystals with $51,7 \,mm$ radius of curvature, with a best resolution of $E / \Delta E \simeq 2000$
obtained for a 008 reflection of $Cu(K_{\alpha})$ lines. As last examples, L. Anklamm et al \cite{HAP3} and H. Legall et al \cite{HAP4} reported,
for measurements exploiting HAPG crystals, again a spectral resolution of $E/\Delta E\simeq \,1800$ in the first reflection order of Ti and Cu, the first one 
exploring also the performances of a fully cilindrical Von Hamos geometry. 
\\All these works, and several others available in literature, report measurements
done in such conditions to have an effective source dimensions of some tens of microns; this configuration is achieved either with microfocused X-ray tubes, 
or with a set of slits and collimators placed before the target to minimize the activated area. Unfortunately, this conditions do not apply in those cases in which,
like the exotic atom X-rays transition, photons are emitted isotropically from a wide and diffused source \cite{SIDD}. It is then very important to test
Von Hamos spectrometers under such conditions, aiming at an optimization between the target size and the minimum acceptable energy resolution demanded by the
single application.
\\For wider sources, the Johann configuration has always been considered more suitable and some 
attempts to use it with silicon or (non mosaic) graphite crystals has been carried out \cite{BEER}; 
however, this configuration has several drawbacks with respect to the Von Hamos one when is to be used in high background environments,
where long distances between the source and the detector are necessary, or when employed to measure lineshapes.
\\In order to allow a higher dynamic range of the spectrometer, the position detector can be rotated such as to have the reflected beam 
impinging perpendicularly on its surface \cite{SEMI}, in the configuration to which, for simplicity, 
from now on we refer to as ``semi''- von Hamos configuration. 

\section{Beam divergence and source width optimization}
\label{sec:bsopt}

\begin{figure}[htbp]
\centering 
\includegraphics[width=7.5cm]{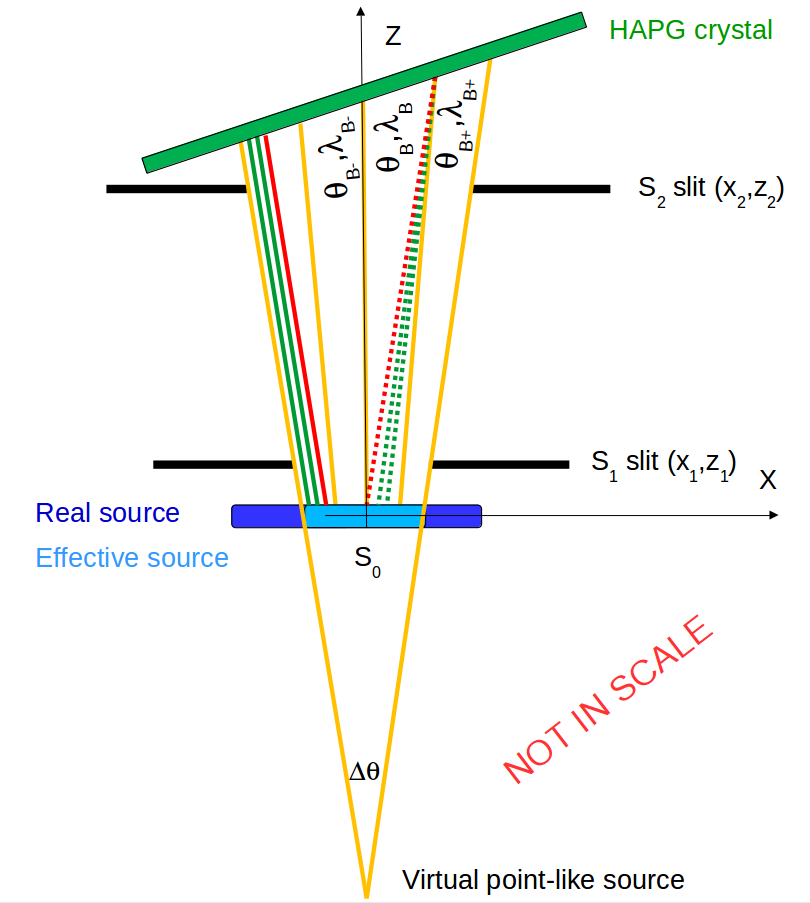}
\qquad
\includegraphics[width=6.5cm]{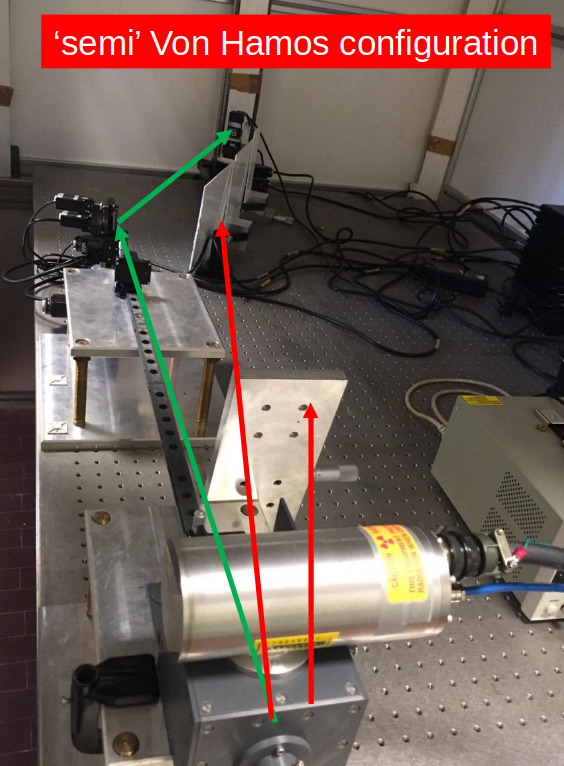}
\caption{\label{setup} \em Left: beam optimization scheme (not in scale); two slits S1 and S2 are used to shape the X-ray beam divergence ($\Delta\theta$) and effective source size ($S_0$). 
Depending on the HAPG mosaic spread, photons emitted parallel (solid lines) and not (dashed lines) to the nominal one matching the Bragg condition (yellow, see text for details) are shown; 
some of them are reflected under the signal peak (solid and dashed green), some are a source of background (solid red), some are not reflected (dashed red). 
See the text for more details. 
Right: picture of the setup installed in the LNF-INFN laboratory.}
\end{figure}

The main scope of the VOXES project is to test the possibility to use the Von Hamos spectrometer, based on HAPG mosaic crystals, to measure 
photons isotropically emitted from an extended (non point-like) source; despite Bragg spectroscopy is a well established technique for high resolution 
X-ray measurements, its main limitations arise in this case. In particular, a dramatic drop in efficiency, with respect to standard solid state X-ray 
detectors, is the most evident one; however, if precisions less then $1\,eV$ are needed, Bragg spectroscopy is still one of the few valid options. 
\\The precision of the energy measurement is a function, mainly, of the event rate and of the peak resolution; depending on the application, 
one would like to have the maximum possible number of signal events keeping the resolution better than a desired value, for example the one needed to distinguish between 
the two $Cu(K_{\alpha_{1,2}})$ lines as in this work, where the performance of the spectrometer is investigated as a function of the beam divergence ($\Delta\theta$) 
and effective source size ($S_0$), in a situation like the one shown in Fig.\ref{setup}.

\noindent As shown in Fig.\ref{setup}, two slits S1 and S2 are used to collimate the characteristic X-ray photons emitted from a Cu foil (the real source in dark blue), 
in order to form a virtual point-like source behind the target instead of a real point-like one on it. Thanks to this configuration, photons emitted from a wider portion of the Cu foil 
(the effective source, $S_0$, in light blue) can be efficiently reflected by the HAPG crystal (green). The two slits define also the divergence of the X-ray beam 
impinging on the crystal ($\Delta\theta$). The yellow lines on the figure represents the photons which, matching the Bragg condition ($\theta_B,\,\lambda_B$), 
form the signal peak on the reflected Bragg spectrum (we call them nominal from now on).
\\Since the photons are isotropically emitted from the whole Cu foil, some of them may have the correct energy and angle to be reflected but originate from a point of the 
target near the nominal one (on the yellow line). 
As far as this mislocation is below the limit given by the mosaic spread of the crystal, such photons are also reflected under the 
signal peak worsening the spectral resolution (solid green lines in the figure); 
on the contrary, when this mislocation exceeds this limit (solid red line) these photons are reflected outside the signal peak.
In the same way, photons not emitted in parallel to the nominal ones may still impinge on the HAPG crystals with an angle below its mosaic spread (dashed green lines) and be then 
reflected under the signal peak also affecting the spectral resolution. On the contrary, if the impinging angle is out of this limit, 
those photons are not reflected on the position detector.
\\The combined effect of $\Delta\theta$ and $S_0$ affects the final resolution and the event rate and is the subject of the measurements reported in this paper.

\section{Voxes setup at LNF}
\label{sec:setup}

The measurements presented in this work have been performed at the INFN Laboratories of Frascati (LNF); the setup used to perform the measurements is shown in Fig.\ref{setup}, right.
\\A $50\,\times 75\,mm$, $2 \,mm$ thick Cu foil (target) is placed inside a 3D printed plastic box and activated by a XTF-5011 Tungsten anode X-ray tube, produced by OXFORD INSTRUMENTS, 
placed on top of the plastic box; the center of the Cu foil represents the origin of the reference frame in which Z is the direction of the characteristic photons emitted by 
the target and forming, with the X axis, the Bragg reflection plane, while Y is the vertical direction, along which primary photons generated by the tube are shot.
The X-ray tube, operated at 20 kV, emits photons from its anode placed 108 mm above the Cu foil; the $127\,\mu m$ thick Be window of the tube allows an angular divergence of $22^{\circ}$.
The Cu target is rotated with respect to the X axis by 45 degrees; this configuration leads to an activated ellipsoidal area of $30\,cm^2$, with $\rho_1=21\,mm$ and $\rho_2=37\,mm$ focuses 
along the foil surface, representing the real source. 
\\Two adjustable slits (Thorlab VA100/M), placed $100 \,mm$ and $635 \,mm$ along the Z axis, respectively, are used to set angular spread of the $Cu(K_{\alpha})$ X-rays beam and 
the transversal dimensions of the effective source.
The vertical dimension of the effective source is instead determined to be $6\,mm$ by the $13,6\,mm$ circular frame of each slit and the $5\,mm$ diameter hole present in the front panel of the plastic box, 
placed at $53\,mm$ along the X axis.
\\The reflecting crystal, produced by the Optigraph GmbH (Berlin, Germany) is cylindrically shaped with a radius of curvature of $206,7\,mm$ and dimensions of $30\,mm\times32\,mm$; 
the substrate is a concave N-BK7 glass lens (Thorlab LK1487L1), coated with a single $100\,\mu m$ HAPG layer with a declared mosaicity of $0.007^{\circ}-0.01^{\circ}$.
\\The position detector is a commercial MYTHEN2-1D 640 channels strip detector produced by Dectris (Zurich, Switzerland); the active area is $32\times8\,mm^2$,
 strip width and thickness are, respectively, $50\,\mu m$ and $420\,\mu m$. The calibration is done using the two known energies of Cu $K_{\alpha 1}$ and $K_{\alpha 2}$.
To avoid direct shining of the primary X-rays on the MYTHEN2-1D surface, an aluminum structure and three $2\,mm$ thick PVC foils have been used, as shown in Fig.\ref{setup} \cite{VOX1}.
\\The spectrometer geometry is optimized for the measurement of the the two Cu $K_{\alpha}$ lines; the corresponding Bragg angles are, for n=1 reflection order, $13,28^{\circ}$ and $13,31^{\circ}$. 
According to the Von Hamos configuration, both source-to-crystal and crystal-to-detector distances are set to $900,54\,mm$ in the XZ dispersion plane.
Given a conversion factor, for the reflected spectrum, of $\simeq \,4,3 \,eV/mm$, the total length of $32\, mm$ of the position detector allows a potential dynamic range of $\simeq 140\, eV$; 
these values are increased to almost $19\, eV/mm$ and $600\, eV$ by rotating the MYTHEN2-1D detector in the ``semi''- von Hamos configuration. 
The achievable energy range is, then, $7750\div8350\, eV$, corresponding to Bragg angles of $13,8^{\circ}$ and $12,8^{\circ}$ respectively. 
Even if such a wide dynamic range is not needed to measure only the Cu $K_{\alpha}$ lines, having the Bragg reflected photons impinging perpendicularly to the MYTHEN2-1D surface enhances 
the detection efficiency with respect to the case in which the same photons, impinging with $\theta_B$ angle on the surface, pass through a thinner portion of the $420\,\mu m$ thick pixels.

\section{Results}
\label{sec:results}

\begin{figure}[htb]
\centerline{%
\includegraphics[width=13.cm]{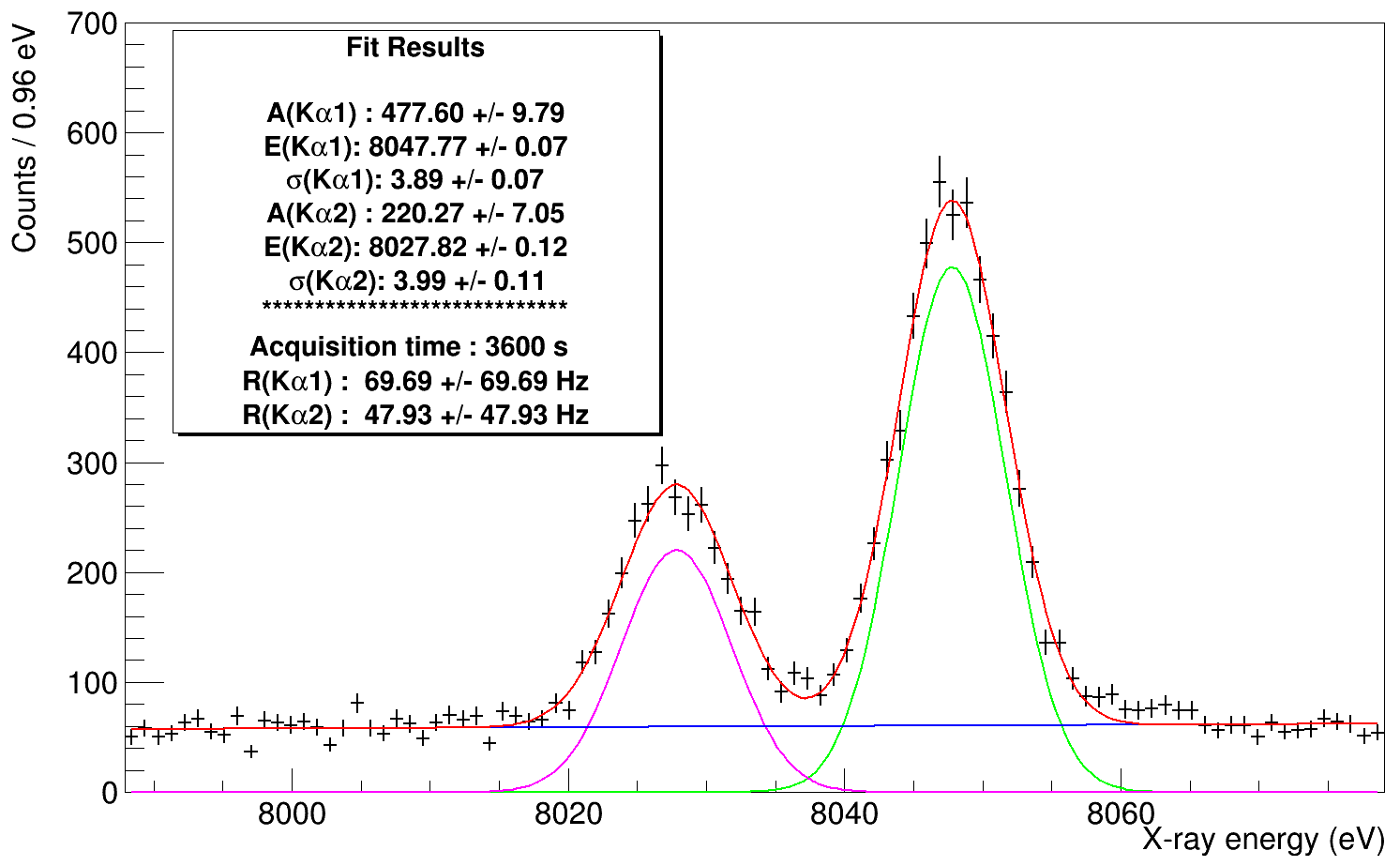}}
\caption{\em Fitted spectrum for $S_0 = 100\,\mu m$ and $\Delta\theta = 0.2^{\circ}$; the spectrum is fitted with two gaussians for the $K_{\alpha 1}$ (green) and $K_{\alpha 2}$ (violet) 
peaks while the background is described with a 2nd order polynomial function (blue). The overall fitting function is shown in red.}
\label{res-spec}
\end{figure}

\begin{figure}[htb]
\centerline{%
\includegraphics[width=16.cm]{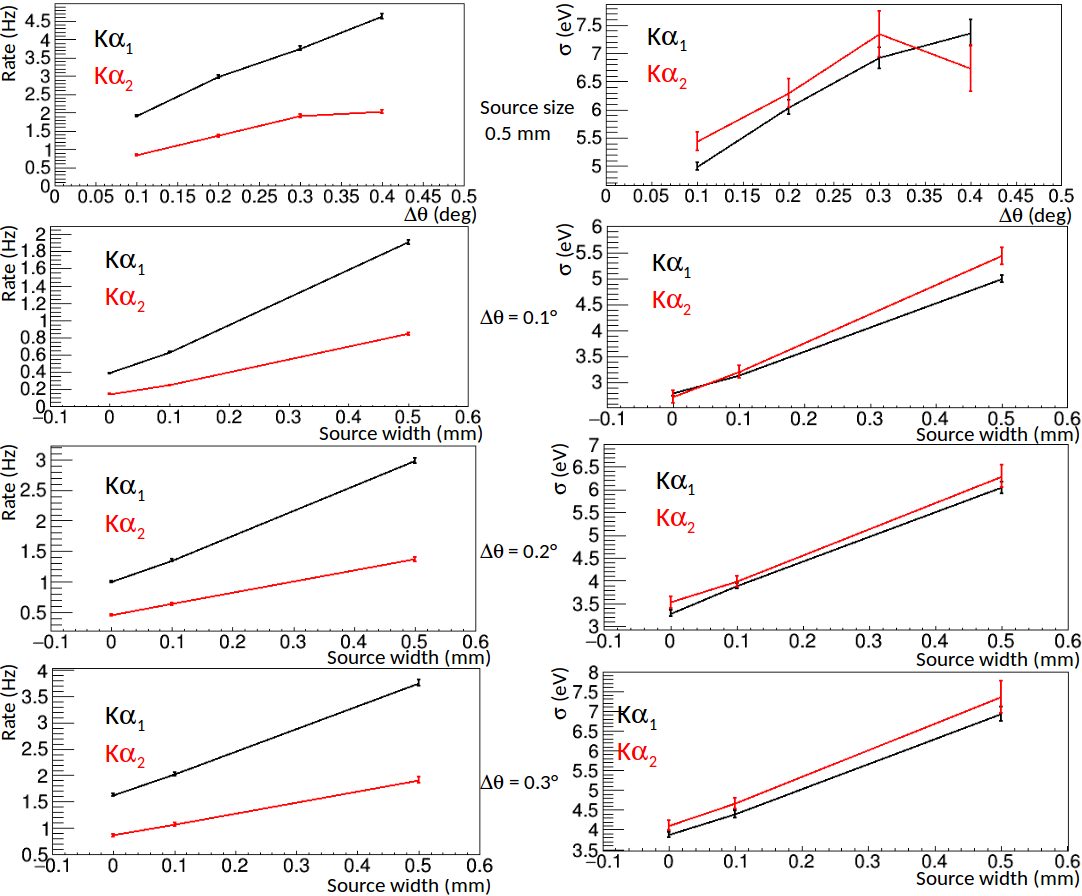}}
\caption{\em Beam divergence ($\Delta\theta$) and source size ($S_0$) dependence plots for $K_{\alpha1}$ (black) and $K_{\alpha2}$ (red) peak; the beam divergence dependence has been studied 
fixing the source size to $0,5\, mm$, for $\Delta\theta$ values of $0,1^{\circ},\,0,2^{\circ},\,0,3^{\circ},\,0,4^{\circ}$, 
while the response of the spectrometer as a function of the source size has been measured for 3 points ($0$, $100\,\mu m$, $500\,\mu m$) for 3 different fixed $\Delta\theta$ ($0,1^{\circ},\,0,2^{\circ},\,0,3^{\circ}$).}
\label{res-trend}
\end{figure}

The Bragg spectra for different values of beam divergence and source width have been acquired in order to investigate the dependence of the signal rate and of the peak resolution on these parameters.
The spectra have been fitted using two gaussian functions for the Cu lines plus a 2nd order polynomial for the background. 
As an example, the fitted spectrum corresponding to $S_0=100\,\mu m$ and $\Delta\theta = 0,2^{\circ}$ (1 hour data taking), is shown in Fig.\ref{res-spec}, while
the signal rate and the peak resolution plots are presented in Fig.\ref{res-trend} for both $K_{\alpha1}$ (black) and $K_{\alpha2}$ (red) lines. 

\noindent The best resolutions ($\sigma$), obtained for $S_0=0\,\mu m$ and $\Delta\theta = 0,1^{\circ}$ in 5 hours of data taking, are $2,81\,eV$ ($E / \Delta E \simeq 1220$) 
and $2,74\,eV$ ($E / \Delta E \simeq 1250$) for the $K_{\alpha1}$ and $K_{\alpha2}$ respectively; 
the $0,05\,eV$ and $0,1\,eV$ obtained precisions dependend on the statistics and scale as $\sigma / \sqrt N$, where N is the number of events under the peak.
\\The beam divergence dependence has been studied, fixing the source size to $0,5\, mm$, for $\Delta\theta$ values of $0,1^{\circ},\,0,2^{\circ},\,0,3^{\circ},\,0,4^{\circ}$, 
while the response of the spectrometer as a function of the source size has been measured for 3 points ($0$, $100\,\mu m$, $500\,\mu m$) 
for 3 different fixed $\Delta\theta$ ($0,1^{\circ},\,0,2^{\circ},\,0,3^{\circ}$).
In Fig.\ref{res-trend}, the strange behaviour for $\Delta\theta=0,4^{\circ}$ in the $K_{\alpha_2}$ (red) resolution (top left corner) is due to a stronger merging of the two peaks, 
caused by the large angle, resulting in a less robust fit of the two lines. 
\section{Conclusions}
\label{sec:conc}

In the framework of the VOXES project, a first set of measurements aiming to investigate the possibility to use the Von Hamos spectrometer with
diffused X-ray source has been carried out. Signal rates and peak resolutions of $Cu(K_{\alpha 1})$ and $Cu(K_{\alpha 2})$ X-ray lines 
have been reported, showing very promising resolutions going from  $2,81\, eV$ to $7,37\, eV \,(\sigma)$ for $K_{\alpha 1}$ 
and from $2,74\, eV$ to $7,35\, eV \, (\sigma)$ for $K_{\alpha 2}$. 
\\These results will be improved, thanks to a more stable and precise positioning system which will be produced at the Stefan-Meyer-Institut f\"ur subatomare Physik in Vienna;
new measurements with a set of crystals having different radius of curvature and mosaicity will be also carried out in 2018, together with an efficiency estimation of the spectrometer.
Finally, the VOXES spectrometer will be tested on the $\pi E-1$ beam line at the Paul Scherrer Institute where, thanks to a $100\div400\,MeV/c$ tuneable momentum pion beam, 
the $4f\rightarrow3d$ and $4d\rightarrow3p$ transition lines of pionic Carbon will be measured \cite{TES2} and the results will be compared to the one obtained,
for the same $\pi C$ lines, with TES detectors by S. Okada et al in 2016 \cite{TES2}.


\acknowledgments

This work is supported by the 5th National Scientific Committee of INFN in the framework of the Young Researcher Grant 2015, n. 17367/2015.
We thank to the LNF staff and in particular to the SPCM service for the support in the preparation of the setup.

\end{document}